**Constraints for non-zero secondary loadings in confirmatory factor analysis**

André Beauducel[1]

Institute of Psychology, University of Bonn, Germany

**Abstract**

In the context of confirmatory factor analysis, the independent clusters model has been found to be overly restrictive in several research contexts. Therefore, a less restrictive criterion for parsimony of non-salient loadings in confirmatory factor analysis was proposed. The criterion is based on 'buffered scales', which have been introduced by Cattell and Tsujioka (1964) as optimal indicators of corresponding factors. Variables with positive and negative loadings on an unwanted factor are balanced out in a buffered scale, so that the variance of the unwanted factor is at minimum. It is proposed here to specify a balance of positive and negative secondary loadings by means of model constraints in order to achieve parsimony of loading patterns. The specification of buffered simple structure by means of model constraints was illustrated by means of a simulation study and an empirical example.

[1] Institute of Psychology, University of Bonn, Kaiser-Karl-Ring 9, 53111 Bonn, Germany, Email: beauducel@uni-bonn.de

## 1 Introduction

In confirmatory factor analysis (CFA), the salient factor loadings are typically freely estimated, whereas the loadings of the variables that are not expected to load on the respective factors are often fixed to zero. In the following, the largest absolute loadings that are used for the interpretation of the factors are termed 'salient' loadings, whereas those absolute loadings that are considerably smaller than the salient loadings and that are usually not used for the interpretation of the factors are termed 'non-salient' or 'secondary' loadings. In several applications of CFA, the variables are expected to have only one salient loading, which corresponds to simple structure. In the following, a model that is based on variables with only one salient loading and with all non-salient loadings being fixed to zero will be termed independent clusters model (ICM). The fixation of non-salient loadings to be zero in the ICM has been criticized because the specification of zero-loadings is often more strict than should be expected from theory (Hofstee, de Raad, & Goldberg, 1992; Vassend & Skrondal, 1997). The overly restrictive specification of zero-loadings leads to subsequent specification searches that may capitalize on chance (MacCallum, Roznowski, & Necowitz, 1992). The combination of the ICM with specification searches might be less convincing than allowing for small non-salient loadings in initial model estimation (Marsh et al., 2009; Marsh, Lüdtke, Nagengast, Morin & Von Davier, 2013), which is possible with exploratory structural equation modeling (ESEM; Asparouhov & Muthén, 2009). However, this introduces the choice of an optimal method of factor rotation into the analysis, whereas the ICM only relies on explicit hypotheses on the variables that load on a factor. Moreover, the ESEM approach does not allow for testing of specific hypotheses on loading structures, because all rotational position of factor axes are equivalent from the perspective of model fit. However, testing hypotheses on loadings was an interesting advance of CFA over exploratory factor analysis.

Another approach that allows to overcome some limitations of the ICM is Bayesian structural equation modeling (BSEM; Muthén & Asparouhov, 2012). In BSEM a prior loading with a small variance and a zero mean can be expected instead of a zero loading (without variance). In contrast to ESEM, BSEM is based on testing hypotheses on factor loadings, but it avoids overly restrictive model specification. However, it might also occur that BSEM obscures the misspecification of parameters (Muthén & Asparouhov, 2012). Further experience with BSEM is needed in order to provide solid guidelines for applied research. Although the BSEM approach allows for a more realistic statistical estimation of the ICM, it does not offer a solution for the conceptual problem that substantial non-salient loadings might occur in the population. When such substantial loadings occur for an ICM, the BSEM approach will also call for specification searches.

To summarize, there are two approaches that allow to overcome some limitations of the ICM with successive specification searches: The ESEM approach, which implies that no hypotheses on specific loading patterns can be tested and that exploratory factor rotation has to be performed. The BSEM approach, which allows for a statistically less restrictive testing of the ICM, although it does not address the problem of substantial non-salient loadings in the population.

Accordingly, the aim of the present paper is to propose an approach that allows to take into account substantial non-salient population loadings (like ESEM) but that remains in the context of hypothesis testing (like BSEM).

Cattell and Radcliffe (1962) and Cattell and Tsujioka (1964) proposed what they called 'buffered scales'. Buffered scales are based on the idea that a set of items with positive loadings on an intended factor might have positive and negative loadings on an unwanted factor. An example of two items with positive loadings on the wanted factor and with a balanced set of positive and negative loadings on the unwanted factor is given in Figure 1 A.

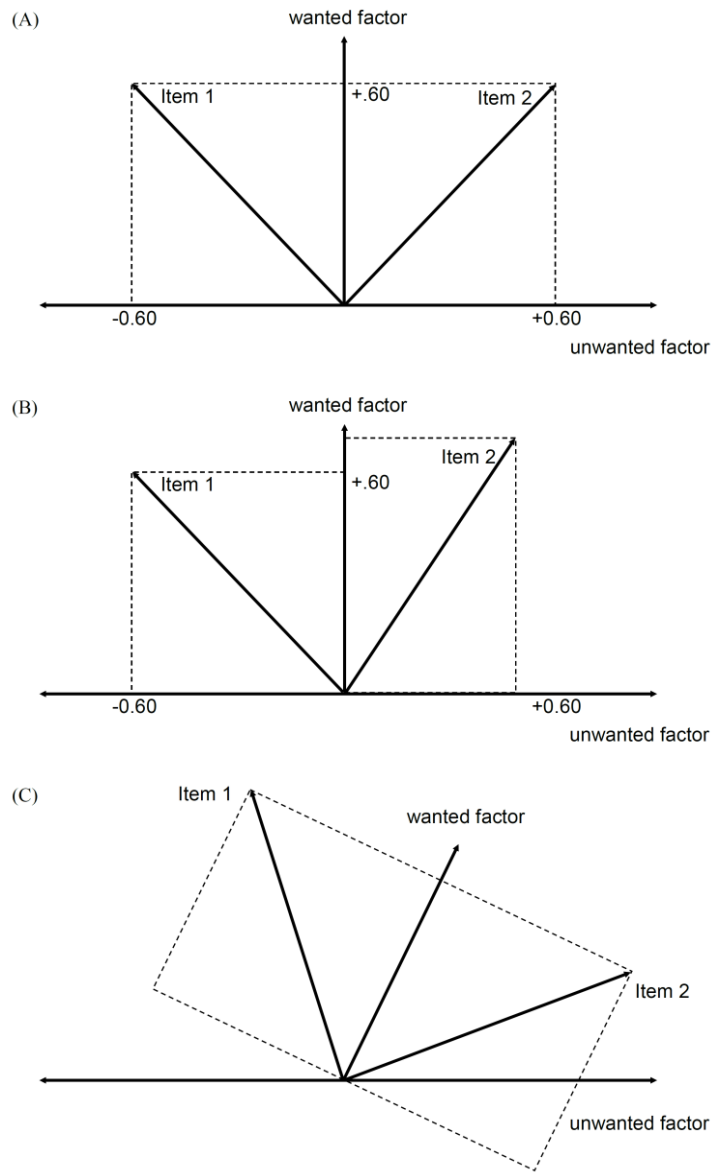

**Figure 1.** Example with two items loading positively on the wanted factor and loading with reversed sign and the same magnitude on the wanted and unwanted factor (A), with a different magnitude on the wanted and unwanted factor (B), and with same magnitude on the wanted and the oblique unwanted factor (C)

Obviously, and in contrast to simple structure, the absolute size of the loadings on the unwanted factor is rather large whereas the sign of the loadings on the unwanted factor is reversed. It follows from this that the absolute size of the item loadings on the unwanted factors can be large as long as the positive and negative item loadings are rather balanced. More formally, one might expect for the first factor that

$$0 = \sum_{i=1}^{k} l_{i1}, \tag{1}$$

where the subscript "1" indicates that the first column of the factor loading matrix is considered, $k$ is the number of non-salient loadings in the first column of the loading matrix, and the subscript "$i$" indicates the respective number of the non-salient loading. An equation is defined for each of the other factors similarly. However, it should also be considered that the items have salient loadings of different magnitude on the wanted factor (see Figure 1 B). A substantial non-salient loading on a wanted factor could have more effect on the estimation of the wanted factor for a variable that loads more on the unwanted factor than for a variable with a smaller loading on the unwanted factor. Therefore, finding the balance of positive and negative non-salient loadings on the wanted factor should take the loading size on the unwanted factor into account. Accordingly, an optimal suppression of unwanted variance could be described as

$$0 = \sum_{i=1}^{k} l_{i1} l_{i2}, \tag{2}$$

where $l_{i1}$ are the non-salient loadings on the first (wanted) factor, whereas $l_{i2}$ describes the salient loadings on the second (unwanted) factor, which are used as the weights for the non-salient loadings of the first factor. A similar definition is considered for each pair of factors. Although Equation 2 can be true when all non-salient loadings are zero as would be expected according to simple structure or ICM, it can also be true when the absolute magnitude of the non-salient loadings is substantial while the signs of the non-salient loadings are balanced so that the weighted sum of the non-salient loadings is zero. Accordingly, a loading pattern that would be compatible with Equation 2 can be similar, but can also be rather different from a loading pattern corresponding to conventional simple structure. Therefore, the right side of Equation 2 comprises simple structure as well as the ICM as a special case.

Although the examples in Figure 1 A and B were based on orthogonal factors, a balanced pattern of non-salient loadings can also occur in oblique factor solutions (Figure 1 C). It can be seen from Figure 1 C that a balanced pattern of positive and negative non-salient loadings can be achieved for oblique factor solutions even when a balanced pattern of non-salient loadings cannot be achieved in the corresponding orthogonal solution. A balanced pattern of positive and negative non-salient loadings will be termed a 'buffered simple structure' in the following.

A balance of positive and negative non-salient loadings implies that the variance of an unwanted factor is balanced out in order to be minimized. Defining models with non-salient loadings with reversed signs may turn out to be promising, especially, when there is a considerable number of substantial non-salient loadings, so that it is impossible to fix all of them to zero in CFA. Thus, buffered simple structure might be regarded as an alternative to the ICM with successive specification searches. For example, specification searches might start from an ICM with secondary/non-salient loadings that are added according to substantial modification indices. This procedure might be justified because the ICM is probably rarely a realistic model. However, this procedure implies that salient loadings are determined according to theory and that non-salient loadings result from exploratory analysis and inductive search. In contrast, specification of buffered simple structure in CFA imposes an a priori theoretical assumption on the non-salient loadings, namely, that their weighted sum is zero. In this sense, the a priori specification of a buffered simple structure might be regarded as more compatible with CFA than a confirmatory ICM combined with exploratory specification search.

Moreover, it seems rather unlikely that selecting items from a domain implies that all items are unrelated with the other domains in the population of items. Especially, in areas where a large number of factors can be found it is extremely unlikely that the items with a salient loading on one factor are completely uncorrelated to all other factors, even when the factors are uncorrelated in the population. Consider again the example of two uncorrelated factors (Figure 1 A, B): It is much more likely to find or select items with small positive and negative population loadings on the unwanted factor than to find or select items with population loadings that are exactly zero, even when the item selection is not performed as a random selection from an item universe. Therefore, the ICM is probably not a realistic model in several domains of research. Whereas the ESEM approach does not contain testable hypotheses on the loading pattern, the buffered simple structure replaces the unrealistic ICM hypothesis of having zero non-salient loadings by the weak but more realistic hypothesis of having a balanced set of positive and negative non-salient loadings. The constraints that are to be imposed in order to get a buffered simple structure imply a testable hypothesis, when the interfactor-correlations are fixed. In this sense, buffered simple structure implies a rather weak hypothesis, but it is considered here that it might be interesting in several areas of research to test a weak hypothesis instead of giving up to test any hypothesis on loading patterns.

Regarding the interpretation of factors with a balanced set of positive and negative non-salient loadings it should be noted that the effect of one factor on the items defining another factor cancel out. Thus, if the items are aggregated, the positive and negative correlations with the unwanted factor will cancel out, so that simple structure is achieved from the perspective of the average of an item set. Therefore, the interpretation of a factor should acknowledge that there might be positive and negative effects of the factor on the items defining another factor, but that these effects are irrelevant when the item set as a whole is considered. Therefore, the interpretation of the factors is similar to the interpretation of factors with conventional simple structure, only that the simple structure applies to the aggregated set of items with non-salient loadings, but not on the single items. One reason for positive and negative effects of a factor on a set of items defining another factor could be that the observed variables (items) are regularly based on language. For example, Vassend and Skrondal (1997) argue that language causes complex logical-semantical relationships between items in the domain of personality. The logical-semantical relationships imply that it is rather unlikely that an item has a single salient loading on one factor and nearly zero loadings on all other factors. It is much more likely that positive as well as negative non-zero loadings occur. If, however, there is a personality factor that causes substantial correlations between some items, this factor will cause the substantial salient loadings. If the additional complex logical-semantical relationships between the items are randomly distributed around the overall effect of the personality factor, they may lead to positive and negative non-salient loadings that cancel out. Thus, even when the population of individuals is considered, the items may have a random distribution of semantical effects. The distribution of loadings may then be the result of a sampling error for the sample of items, not for the sample of individuals. Similar ideas may be developed for other domains of research.

The aim of the present study is therefore to propose and explore the use of the concept of balanced non-salient loadings as a basis for buffered simple structure as an alternative to conventional simple structure. In order to achieve this, first, a more complete description of buffered simple structure is presented. Then, methods for defining buffered simple structure by means of CFA are proposed. The methods are illustrated by means of a simulated population example. In order to compare the ICM with buffered simple structure a small simulation study was performed. Finally, an empirical example is presented in order to compare the results based on methods that allow for the identification of buffered simple structure with the ICM.

## 1.1 Buffered simple structure

It should be noted that the specification of the non-salient loadings according to the criterion expressed in Equation 2 implies for a $p$ variables x $q$ factors loading matrix $\mathbf{L}$ that $\mathbf{L'L}$ is a diagonal matrix for a two-factor model. In a two-factor model the variables with non-salient loadings on the first factor have salient loadings on the second factor (and vice versa) unless the absolute loadings on the two factors are equal. It is therefore possible to define buffered simple structure as

$$\mathbf{0} = \mathbf{1}'\big(\mathbf{L'L} - diag\big(\mathbf{L'L}\big)\big)\mathbf{1}, \text{ for } q \leq 2, \tag{3}$$

where $\mathbf{1}$ is a unit column vector of order $q$ x 1 and $\mathbf{L}$ the pattern of factor loadings. Equation 3 yields zero if $\mathbf{L'L} = diag\ (\mathbf{L'L})$. A balanced set of positive and negative non-salient loadings implies that the weighted sum of the non-salient loadings corresponding to each subset of variables with salient loadings on one factor is a minimum. However, for $q > 2$ buffered simple structure does not imply $\mathbf{L'L} = diag\ (\mathbf{L'L})$ for the columns of non-salient loadings. An example for a three-factor loading pattern $\mathbf{L}$ composed by three submatrices or blocks $\mathbf{B}_i$, each comprising a column $\mathbf{S}_i$ of three salient loadings and two columns $\mathbf{N}_{ij}$ of three non-salient loadings, is

$$
\mathbf{L} = \begin{bmatrix} \mathbf{B}_1 \\ \mathbf{B}_2 \\ \mathbf{B}_3 \end{bmatrix} = \begin{bmatrix} \mathbf{S}_1 & \mathbf{N}_{11} & \mathbf{N}_{12} \\ \mathbf{N}_{21} & \mathbf{S}_2 & \mathbf{N}_{22} \\ \mathbf{N}_{31} & \mathbf{N}_{32} & \mathbf{S}_3 \end{bmatrix} = \begin{bmatrix} s_{11} & n_{12} & n_{13} \\ s_{21} & n_{22} & n_{23} \\ s_{31} & n_{32} & n_{33} \\ n_{41} & s_{42} & n_{43} \\ n_{51} & s_{52} & n_{53} \\ n_{61} & s_{62} & n_{63} \\ n_{71} & n_{72} & s_{73} \\ n_{81} & n_{82} & s_{83} \\ n_{91} & n_{92} & s_{93} \end{bmatrix} . \tag{4}
$$

In order to achieve a buffered simple structure with respect to each factor, it is necessary to define the following six constraints for the loading blocks of the matrix presented in Equation 4 as

$$
\begin{aligned}
0 &= tr(\mathbf{S}_1' \mathbf{N}_{11}), 0 = tr(\mathbf{S}_1' \mathbf{N}_{12}), \\
0 &= tr(\mathbf{S}_2' \mathbf{N}_{21}), 0 = tr(\mathbf{S}_2' \mathbf{N}_{22}), \\
0 &= tr(\mathbf{S}_3' \mathbf{N}_{31}), 0 = tr(\mathbf{S}_3' \mathbf{N}_{32}),
\end{aligned} \tag{5}
$$

where "tr" denotes the trace. The subscripts in the constraints are based on the matrix in the middle of Equation 4. For a larger number of factors, the constraints can be specified similarly. Even when the number of salient loadings and non-salient loadings is not the same for each block, a constraint based on the sum of weighted loadings can be specified.

According to Jöreskog (1969) an unrestricted solution, corresponding to an exploratory factor solution, is obtained when the number of fixed parameters equals $q^2$. With six constraints for the loading matrix and three fixed factor variances, the model corresponds to an unrestricted solution for $q = 3$. It is reasonable to specify a constraint for each block of salient and non-salient loadings, so that the number of constraints is sufficient for model identification. However, with $q$ - 1 constraints for each column of the loading matrix, the orthogonal model with fixed factor variances is unique (Jöreskog, 1969) and represents a testable hypothesis.

It should, however, be noted that the constraints can also become zero if the salient loadings $\mathbf{S}_i$ are zero. It is therefore necessary to provide a method that excludes that the absolute size of $\mathbf{S}_i$ is minimized. There are two different ways to ascertain that the buffered simple structure is based on a sufficient number of substantial salient loadings. One way is to start from an ICM with substantial salient loadings, to use the salient loadings of that model and to insert them as fixed values into the constraints. From this perspective, estimating buffered simple structure can be regarded as a method for relaxing the specification of non-salient loadings of a poor fitting ICM without relying on model modification indices (see below for a description of this procedure).

Another possibility is to provide a condition that excludes a minimization of the absolute size of the salient loadings for each block of loadings by means of defining a matrix

$$\mathbf{H}_i = \mathbf{S}_i^{\circ 2} + 1 \tag{6}$$

where "$^{\circ 2}$" indicates the element-wise power. Then, the matrix $\mathbf{H}_i$ can be used for computing constraints as

$$\mathbf{0} = \mathrm{tr}\left(\mathbf{H}_i'\mathbf{N}_{ij}\right), \text{for } i = 1 \text{ to } q, j = 1 \text{ to } q - 1. \tag{7}$$

Accordingly, the weights for the salient loadings will be greater one, when is $\mathbf{H}_i$ used as a weight matrix. Obviously, a zero value in Equation 7 can only be obtained by means of a weighted sum of positive and negative non-salient loadings or if the non-salient loadings in $\mathbf{N}_{ij}$ are zero. The number of constraints for each column of the loading matrix is $q - 1$. When constraints corresponding to Equation 7 are used, it is not necessary to start with an initial ICM so that a single modeling step is sufficient for obtaining a buffered simple structure. Therefore, this procedure is termed the 'one-step' procedure in the following. In order to get an identified model, it is, however, necessary to specify a constraint for each combination of salient and non-salient loadings in order to get $q - 1$ constraints for each column of the loading matrix. The 'one-step' procedure leads to a just identified (unrestricted) model when the factor variances are fixed and when interfactor-correlations are freely estimated. It leads to a testable hypothesis on the loadings only for orthogonal models or when the interfactor-correlations are fixed to specific values. If researchers regard the buffered simple structure more as a refinement of an initial ICM, they will probably opt for the successive modeling procedure. In the successive procedure salient-loadings and interfactor-correlations are estimated first. Then, the salient loadings are used in order to specify model constraints as in Equation 5 and the interfactor-correlations are fixed in order to provide a testable hypothesis. Since it is possible to fix the interfactor-correlations to the values obtained in the initial ICM, the successive modeling procedure can more easily be used in order to test for a buffered simple structure with correlated factors.

## 1.2  Finding buffered simple structure

Specifying models with perfect buffered simple structure by means of the model constraints and with fixed interfactor-correlations implies that any departure from buffered simple structure will lead to an increase of model misfit. An example for the one-step modeling based on model constraints corresponding to Equation 7 is provided in the next section. The modeling steps that are necessary when an ICM is used as a starting model for the multi-step procedure are summarized in Table 1. In a first step, the salient loadings are to be estimated in the ICM (Model 1). Alternatively, salient loading estimates may also be taken from an exploratory factor solution. In a second step, the salient loadings of Model 1 are used as weights for the weighted sum of the non-salient loadings for each block of salient loadings. These weighted sums are constrained to be zero in order to specify and estimate the corresponding buffered simple structure model while the salient loadings are again freely estimated (Model 2). If the salient loadings estimated in Model 2 diverge substantially from the salient loadings used as initial weights in the model constraint, the weights in the model constraints should be replaced by the salient loadings of Model 2, and the model should again be estimated (Model 3). This procedure can be repeated until the salient loadings estimated from the model correspond exactly to the salient loadings from the previous model used as weights for the model constraints. Finally, model fit should be evaluated. An advantage of the initial ICM estimation is that the factor inter-correlations can be fixed according to the results of the ICM. This allows to test as a hypothesis whether the buffered-simple structure fits to the data. An example for a corresponding Mplus syntax is presented together with the following population example.

**Table 1.** Multi-step procedure for the estimation of buffered simple structure by means of CFA

| Step (Model) | Salient loadings | Factor inter-correlations | Non-salient loadings | Model constraint |
|---|---|---|---|---|
| Model 1 | Free estimation | Free estimation | Fixed to zero | - |
| Model 2 | Free estimation | Fixed to ICM values | Free estimation | The sum of the non-salient loadings weighted by the salient loadings from Model 1 is zero |
| Model 3 | Free estimation | Fixed to ICM values | Free estimation | The sum of the non-salient loadings weighted by the salient loadings from Model 2 is zero |
| Repeat *i* times until the weights in the model constraint and the salient loadings resulting from free estimation are equal: | | | | |
| Model (3 + *i*) | Free estimation | Fixed to ICM values | Free estimation | The sum of the non-salient loadings weighted by the salient loadings from Model (2 + *i*) is zero |

## 1.3  Population example: buffered simple structure

This loading matrix presented in Table 2 has a perfect buffered simple structure, because the sum of the non-salient loadings multiplied by the salient loadings is zero for each block of salient loadings, i.e., the positive and negative non-salient loadings are perfectly balanced. The population correlation matrix generated from the population model presented in Table 2 was submitted to CFA (Mplus 7.11; Muthén & Muthén, 1998-2013). The first estimation of buffered simple structure was based on the one-step modeling according to Equation 7. The complete model specification (Mplus syntax) containing the corresponding model constraint for each column of the loading matrix can be found in Appendix A. Since no sample size information is relevant for the population data, the fit is only evaluated by means of the standardized root mean squared residual (SRMR). The model fits almost perfectly to the data (SRMR = .002) and the model estimates obtained for this model correspond exactly to the estimates obtained in Model 3 as well as to the population model (see Table 2). In order to illustrate the multi-step procedure, an ICM was estimated by means of maximum likelihood estimation. Salient loadings were freely estimated, non-salient loadings were fixed to zero, factor variances were fixed to one, and factor inter-correlations were freely estimated. The ICM model fits already quite well to the data (SRMR = .073). However, the subsequent modeling is described in order to illustrate how to find a buffered simple structure. The resulting salient loading estimates (Model 1, see Table 2) were used as weights for the model constraints imposed on the secondary loadings when Model 2 was estimated and the factor inter-correlations were fixed according to the initial ICM. The model specification and the model constraints are presented in Appendix B. Model 2 fits almost perfectly to the data (SRMR = .002) and the resulting parameter estimates presented in Table 2 are nearly identical to the parameters of the population model. However, the salient loadings used for the constraints (.594) were not identical to the salient loadings estimated for Model 2 (.600). Therefore, a final model based on the salient loading estimates of Model 2 as weights in the model constraints was calculated (Model 3). However, this final step does not alter the parameter estimates and model fit, so that no additional columns were presented in Table 2. The only difference between Model 2 and Model 3 is that the salient loadings used in the model constraints condition (.600) are identical to the salient loading estimates of Model 3. Thus, three steps were necessary in order to obtain a buffered simple structure by means of an initial ICM (Model 1: ICM, Model 2: model constraints, Model 3: adjusted model constraints). When variable x5 and x6 as well as variable x5 and x10 are switched in the model constraints, the model fits drops considerably (SRMR = .139). Thus, although all loadings are freely estimated, a wrong placement of the loadings in the model constraints can lead to model rejection. However, model rejection is only possible when the factor inter-correlations are fixed, which is more easily be obtained in the multi-step procedure.

In order to illustrate the difference between the buffered simple structure obtained by means of CFA with a more conventional procedure, a specification search starting from Model 1 (ICM) was performed (see Table 2). Modification indices for Model 1 were inspected and for each factor three non-salient loadings with the largest modification indices (> 15) were freely estimated (Model 4). A larger number of freed modification indices would lead to collapsing factors or to identification problems. The resulting Model 4 had an acceptable fit (SRMR = .059).

**Table 2.** CFA-example: Population parameters, parameter estimates for ICM (Model 1), buffered simple structure (Model 2/3), and model resulting from specification search (Model 4)

| Item | Population | | | Model 1 (ICM) | | | Model 2 / Model 3* | | | Model 4 (spec. search) | | |
|---|---|---|---|---|---|---|---|---|---|---|---|---|
| | F1 | F2 | F3 | F1 | F2 | F3 | F1 | F2 | F3 | F1 | F2 | F3 |
| x1 | **.600** | .150 | -.150 | **.595** | - | - | **.600** | .149 | -.149 | **.707** | - | -.261 |
| x2 | **.600** | .150 | -.150 | **.595** | - | - | **.600** | .149 | -.149 | **.707** | - | -.261 |
| x3 | **.600** | .150 | -.150 | **.595** | - | - | **.600** | .149 | -.149 | **.707** | - | -.261 |
| x4 | **.600** | -.150 | .150 | **.595** | - | - | **.600** | -.149 | .149 | **.773** | -.322 | - |
| x5 | **.600** | -.150 | .150 | **.595** | - | - | **.600** | -.149 | .149 | **.773** | -.322 | - |
| x6 | **.600** | -.150 | .150 | **.595** | - | - | **.600** | -.149 | .149 | **.773** | -.322 | - |
| x7 | -.150 | **.600** | .150 | - | **.595** | - | -.149 | **.600** | .149 | -.206 | **.715** | - |
| x8 | -.150 | **.600** | .150 | - | **.595** | - | -.149 | **.600** | .149 | -.206 | **.715** | - |
| x9 | -.150 | **.600** | .150 | - | **.595** | - | -.149 | **.600** | .149 | -.206 | **.715** | - |
| x10 | .150 | **.600** | -.150 | - | **.595** | - | .149 | **.600** | -.149 | - | **.593** | - |
| x11 | .150 | **.600** | -.150 | - | **.595** | - | .149 | **.600** | -.149 | - | **.593** | - |
| x12 | .150 | **.600** | -.150 | - | **.595** | - | .149 | **.600** | -.149 | - | **.593** | - |
| x13 | -.150 | .150 | **.600** | - | - | **.595** | -.149 | .149 | **.600** | - | - | **.593** |
| x14 | -.150 | .150 | **.600** | - | - | **.595** | -.149 | .149 | **.600** | - | - | **.593** |
| x15 | -.150 | .150 | **.600** | - | - | **.595** | -.149 | .149 | **.600** | - | - | **.593** |
| x16 | .150 | -.150 | **.600** | - | - | **.595** | .149 | -.149 | **.600** | - | - | **.597** |
| x17 | .150 | -.150 | **.600** | - | - | **.595** | .149 | -.149 | **.600** | - | - | **.597** |
| x18 | .150 | -.150 | **.600** | - | - | **.595** | .149 | -.149 | **.600** | - | - | **.597** |
| inter-factor correlations | | | | | | | | | | | | |
| | 1.000 | | | 1.000 | | | 1.000 | | | 1.000 | | |
| | .300 | 1.000 | | .304 | 1.000 | | .304 | 1.000 | | .624 | 1.000 | |
| | .300 | .300 | 1.000 | .304 | .304 | 1.000 | .304 | .304 | 1.000 | .477 | .358 | 1.000 |

*Note.* Salient loadings are given in bold face. * The result obtained for Model 2 and Model 3 can also be obtained directly by means of the one-step procedure based on Equation 7 (see Appendix A).

The loading pattern shows quite substantial non-salient loadings, especially on the second factor. Moreover, the loading size of the salient loadings is quite different in Model 4, although the salient loadings were of equal size in Model 1 as well as in Model 2 and 3. This demonstrates that substantial differences in the variables representing the factors are induced by the ICM with subsequent specification search, although these differences were not present in the population model. This shows that the ICM specification search strategy can be misleading when the data correspond to a buffered simple structure.

## 2 Simulation Study

Since Equation 7 is also true for the ICM, a model constraint imposing buffered simple structure should also allow for an ICM if this corresponds to the data. It is, however, not clear whether the salient loadings are estimated with the same precision as with the ICM when a buffered simple structure is specified for data that correspond to the ICM. Does the model constraint for the specification of buffered simple structure lead to a distortion of the salient loading estimates when the non-salient loadings are zero in the population? This question was investigated by means of a simulation study. Since buffered simple structure allows for the ICM as a special case, it is expected that the model constraints for buffered simple structure lead to precise salient loading estimates for population models corresponding to the ICM as well as for population models corresponding to buffered simple structure. Accordingly, the ICM was used as a population model for data generation as well as models with larger secondary loadings that correspond to a buffered simple structure.

Population models with six salient loadings per factor ($p/q$) with three factors ($q = 3$), moderate ($l = .60$) and large ($l = .80$) salient loadings, with uncorrelated and slightly correlated factors (.30), and with a non-salient loading size of .00 were investigated as models with perfect simple structure (ICM). Models with buffered simple structure were generated as follows: For models with $l = .60$, models with an absolute non-salient loading size ($anl$) of .05, as well as models with $anl = .10$, .15, and .20 were generated. The same set of models was generated for $l = .80$. The signs of the non-salient loadings were distributed in a way that results in a perfect buffered simple structure. Thus, the population loading matrix corresponds to the constraints defined in Equation 6. The buffered simple structure models were generated for orthogonal factors as well as for slightly correlated factors.

All models based on $anl = .00$ are ICM and all models based on $anl = .05$, .10, .15, and .20 are buffered simple structures. Each of the (2 salient loading sizes x 5 non-salient loading sizes x 2 degrees of interfactor-correlations=) 20 population models was investigated in 1,000 samples with $n = 150$, 300, and 900 cases. The root mean squared difference (RMSD) between the population loadings and the corresponding estimated loadings was averaged across the 1,000 samples per condition. For the orthogonal population models, the interfactor-correlations were fixed to zero for model estimation, whereas for the population models with interfactor-correlations of .30, the interfactor-correlations were set free for model estimation in the samples.

For all conditions the mean Root Mean Square Error of Approximation (RMSEA) of the estimated buffered simple structure models was below .05, indicating an acceptable model fit. For the estimated ICM the mean RMSEA was greater .05 for all conditions with $anl \geq .10$, indicating that the ICM has a moderate or poor fit under these conditions. The RMSD for the loadings of the three-factor models are presented in Figure 2 (A-D) and the RMSD for the correlations of the models with correlated factors are presented in Figure 3 (A, B).

(A)  $l$ = .60, interfactor-correlation = .00

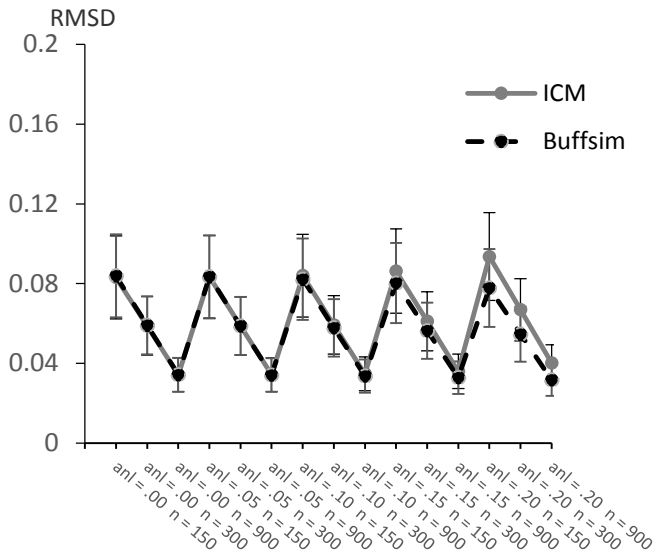

(B)  $l$ = .60, interfactor-correlation = .30

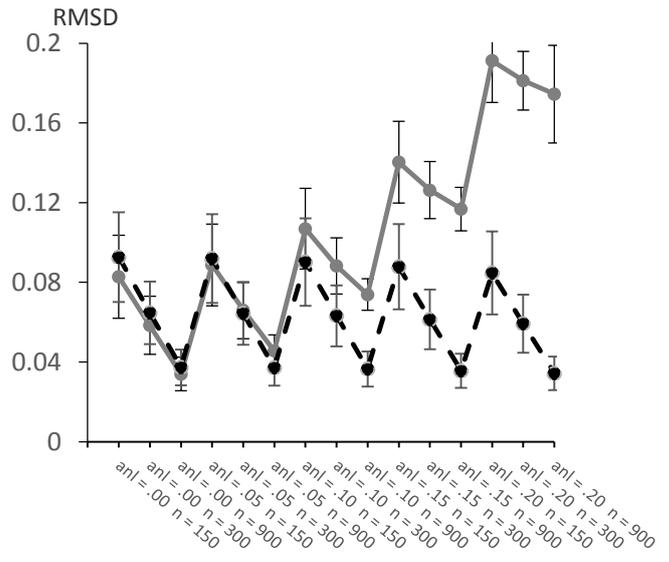

(C)  $l$ = .80, interfactor-correlation = .00

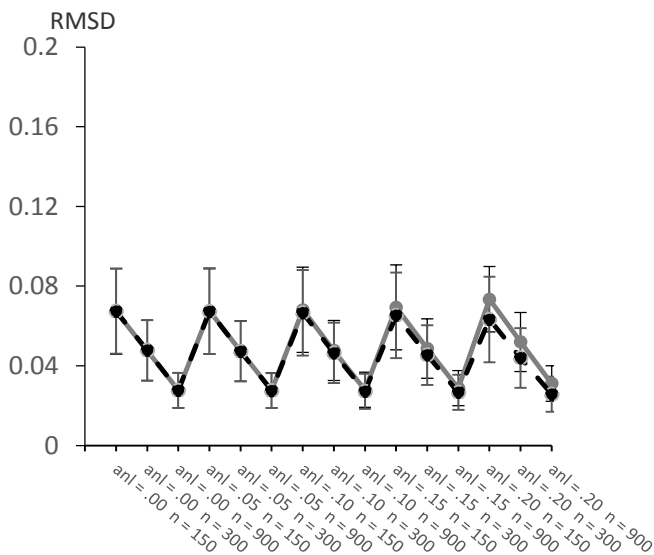

(D)  $l$ = .80, interfactor-correlation = .30

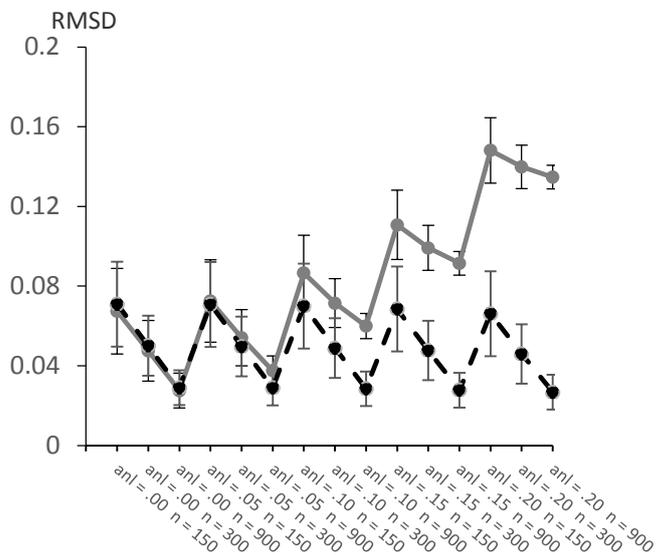

**Figure 2.** Root mean squared difference (RMSD) between population loadings and estimated sample loadings (error bars represent standard errors); $l$ = salient loadings; $anl$ = absolute non-salient loadings; $n$ = sample size

The results for the conditions representing the ICM (with zero absolute non-salient population loadings; $anl$ = .00) and for buffered simple structure ($anl >$ .00) are presented together. When the population model is an ICM (*anl* = .00), the RMSD of the salient loadings was nearly same when an ICM or a buffered simple structure was specified (see Figure 2). However, with increasing *anl* in the population, the RMSD of the salient loadings based on the specification of the ICM increased, whereas the RMSD based on buffered simple structure remained nearly constant. Since the RMSD for *l* = .60 and *n* = 150 and correlated factors was slightly larger for the buffered simple structure than for the ICM, it is recommended to estimate an initial ICM when the sample size is small and when non-zero interfactor-correlations as well as small salient loadings are to be expected.

As for the factor loadings, the precision of reproducing the population interfactor-correlations was assessed by means of the mean squared difference (RMSD) between the population interfactor-correlations and the corresponding interfactor-correlations estimated from the samples (see Figure 3). As for the loadings, no relevant differences between model estimation based on the ICM and model estimation based on buffered simple structure occurred for *anl* = .00, whereas the RMSD of ICM estimation increased with increasing *anl*. The RMSD for model estimation based on buffered simple structure remained nearly constant.

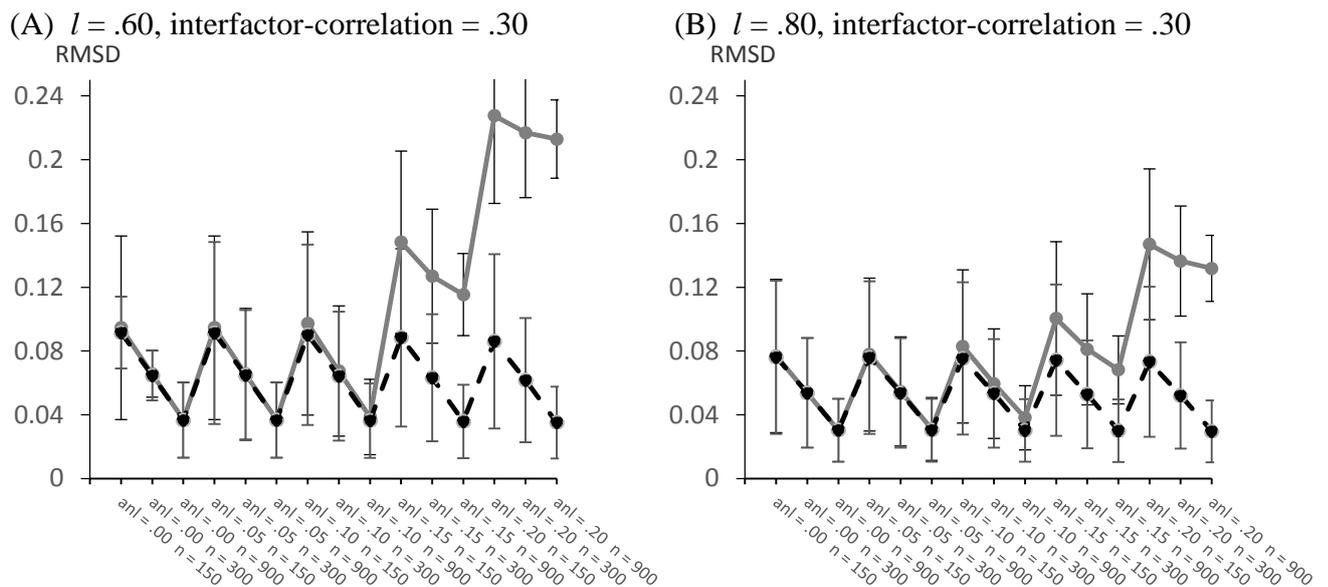

(A) *l* = .60, interfactor-correlation = .30      (B) *l* = .80, interfactor-correlation = .30

Figure 3. Root mean squared difference (RMSD) between population interfactor-correlations and estimated sample interfactor-correlations (error bars represent standard errors); *l* = salient loadings; *anl* = absolute non-salient loadings; *n* = sample size

## 3  Empirical example

The empirical example is based on a sample of 587 German participants (383 Females; Age in years: M = 33.85, SD = 12.40, range: 16-65). The participants were recruited through newspaper advertising, indicated informed consent and filled in the German Version of the NEO-PI-R (Ostendorf & Angleitner, 2004). A test for multivariate normality of the 30 facet scales comprising eight items each was performed with PRELIS 2.80 (Jöreskog & Sörbom, 2006) and revealed that the data do not fit to this assumption ($\chi^2$(2)=467.06, $p < .001$). Therefore, a robust maximum likelihood estimation was performed with Mplus 7.11 and the Satorra-Bentler scaled $\chi^2$-statistic was reported. Moreover, the RMSEA, the Comparative Fit Index (CFI), and the Standardized Root Mean Square Residual (SRMR) were reported in order to evaluate model fit. First, an ICM (Model 1; see Table 3) based on freely estimated salient loadings, non-salient loadings that were fixed to zero, unit-variances of the factors, and freely estimated factor inter-correlations was estimated. Model 1 did not fit to the data ($\chi^2_{SB}$(395)=3430.07, $p < .001$; RMSEA = .114; CFI = .627; SRMR = .143). The salient loadings of Model 1 were used as a starting point for modeling a buffered simple structure by means of CFA. All loadings were set free in the next model (Model 2) factor variances were fixed to one, factor inter-correlations were freely estimated, and model constraints defining the linear combination of the salient loadings with the non-salient loadings to be zero were added for each block of salient loadings. The fit of Model 2 was moderate but acceptable ($\chi^2_{SB}$(295)= 925.17, $p < .001$; RMSEA = .060; CFI = .923; SRMR = .030). However, the salient loadings that are estimated in Model 2 did not correspond exactly to the salient loadings that were obtained from Model 1 (ICM) and that were used in the model constraints. Therefore, Model 2 is a first approximation to buffered simple structure that can already interpreted, but slight further improvements of buffered simple structure were possible when the salient loading estimates correspond exactly to the salient loading estimates used in the model constraints. In order to improve buffered simple structure, the salient loadings estimates of Model 2 were used for the model constraints of Model 3. In turn, the salient loading estimates of Model 3 were used in the model constraints of Model 4, and the final Model 5 was based on the salient loading estimates of Model 4. In Model 5, the salient loading estimates used for the model constraints and the salient loading estimates obtained from model estimation were identical so that no further improvement was possible. However, these adjustments of the salient loading estimates were small and had no effect on model fit, since the fit of Model 5 ($\chi^2_{SB}$(295)= 925.17, $p < .001$; RMSEA = .060; CFI = .923; SRMR = .030) was identical to the fit of Model 2. The loadings and inter-factor correlations for Model 5 were presented in Table 3. The quality of buffered simple structure was calculated according to Equation 6. A result of .076 indicates that the sum of the salient loadings multiplied by the non-salient loadings is rather small across all blocks of salient loadings. Thus, a buffered simple structure was obtained in Model 5.

**Table 3.** Empirical example: Completely standardized solution for ICM (Model 1), buffered simple structure (Model 5), and the model resulting from specification search (Model 6)

| Item | Model 1 (ICM) | | | | | Model 5 | | | | | Model 6 (specification search) | | | | |
|---|---|---|---|---|---|---|---|---|---|---|---|---|---|---|---|
| | N | E | O | A | C | N | E | O | A | C | N | E | O | A | C |
| n1 | **.797** | - | - | - | - | **.925** | .070 | .021 | .053 | .179 | **.934** | .173 | - | - | .203 |
| n2 | **.670** | - | - | - | - | **.812** | .081 | -.072 | -.340 | .062 | **.820** | - | - | -.423 | - |
| n3 | **.869** | - | - | - | - | **.811** | -.096 | .042 | .062 | -.009 | **.858** | - | - | - | - |
| n4 | **.724** | - | - | - | - | **.624** | -.202 | -.008 | .156 | .004 | **.689** | - | - | .157 | - |
| n5 | **.375** | - | - | - | - | **.451** | .449 | .022 | -.191 | -.341 | **.439** | .415 | - | -.364 | -.283 |
| n6 | **.878** | - | - | - | - | **.746** | -.068 | -.041 | .076 | -.186 | **.793** | - | - | - | -.189 |
| e1 | - | **.690** | - | - | - | .096 | **.739** | .068 | .429 | .068 | - | **.849** | - | .270 | .256 |
| e2 | - | **.583** | - | - | - | .133 | **.772** | -.111 | .163 | -.139 | - | **.688** | - | - | - |
| e3 | - | **.531** | - | - | - | -.143 | **.508** | -.009 | -.353 | .216 | -.220 | .339 | - | **-.439** | .253 |
| e4 | - | **.536** | - | - | - | .084 | **.612** | -.064 | -.140 | .247 | - | **.469** | - | -.265 | .321 |
| e5 | - | **.391** | - | - | - | .054 | **.515** | -.054 | -.272 | -.217 | - | .361 | - | **-.411** | -.167 |
| e6 | - | **.819** | - | - | - | -.217 | **.584** | .195 | .113 | -.067 | -.340 | **.478** | .223 | - | - |
| o1 | - | - | **.606** | - | - | .016 | -.012 | **.632** | -.083 | -.302 | - | -.180 | **.712** | - | -.380 |
| o2 | - | - | **.690** | - | - | .176 | -.037 | **.742** | .158 | .166 | - | -.326 | **.985** | .320 | - |
| o3 | - | - | **.749** | - | - | .253 | .247 | **.572** | .113 | .106 | - | - | **.755** | .185 | - |
| o4 | - | - | **.517** | - | - | -.285 | .206 | **.387** | -.035 | -.261 | -.395 | - | **.491** | - | -.312 |
| o5 | - | - | **.497** | - | - | -.127 | -.159 | **.708** | -.170 | .185 | -.254 | -.454 | **.854** | - | - |
| o6 | - | - | **.454** | - | - | -.278 | .054 | **.384** | .192 | -.264 | -.351 | - | **.420** | .248 | -.290 |
| a1 | - | - | - | **.402** | - | -.266 | .281 | .050 | **.488** | -.069 | -.349 | .381 | - | **.450** | - |
| a2 | - | - | - | **.628** | - | -.019 | -.265 | .006 | **.591** | .030 | -.195 | -.099 | - | **.647** | - |
| a3 | - | - | - | **.670** | - | .086 | .317 | .007 | **.729** | .146 | .127 | .483 | - | **.679** | .278 |
| a4 | - | - | - | **.588** | - | -.203 | -.195 | -.002 | **.654** | -.076 | - | - | - | **.725** | - |
| a5 | - | - | - | **.658** | - | .141 | -.142 | -.097 | **.607** | -.041 | - | - | - | **.639** | - |
| a6 | - | - | - | **.616** | - | .228 | .174 | .136 | **.553** | .011 | - | - | .370 | **.588** | - |
| c1 | - | - | - | - | **.715** | -.259 | .129 | .138 | -.127 | **.572** | -.363 | - | .183 | - | **.549** |
| c2 | - | - | - | - | **.640** | .160 | -.044 | -.123 | -.069 | **.737** | .147 | - | -.139 | - | **.741** |
| c3 | - | - | - | - | **.672** | .121 | .033 | -.063 | .229 | **.785** | - | - | - | .295 | **.767** |
| c4 | - | - | - | - | **.691** | .086 | .144 | .170 | -.187 | **.721** | - | - | .233 | -.132 | **.680** |
| c5 | - | - | - | - | **.850** | -.119 | .115 | -.112 | .004 | **.736** | -.196 | - | - | - | **.718** |
| c6 | - | - | - | - | **.471** | -.059 | -.378 | .062 | .138 | **.561** | - | -.261 | - | .270 | **.536** |

factor inter-correlations

| | | | | | | | | | | | | | | | |
|---|---|---|---|---|---|---|---|---|---|---|---|---|---|---|---|
| N | 1.000 | | | | | 1.000 | | | | | 1.000 | | | | |
| E | -.452 | 1.000 | | | | -.366 | 1.000 | | | | -.217 | 1.000 | | | |
| O | -.002 | .605 | 1.000 | | | -.016 | .479 | 1.000 | | | -.055 | .666 | 1.000 | | |
| A | .142 | .006 | .127 | 1.000 | | .081 | -.110 | .025 | 1.000 | | .289 | -.096 | -.247 | 1.000 | |
| C | -.518 | .179 | -.051 | -.076 | 1.000 | -.418 | .068 | -.038 | -.040 | 1.000 | -.361 | -.178 | .034 | -.194 | 1.000 |

*Note.* Salient loadings are given in bold face.

The buffered simple structure was compared with a model (Model 6) that was obtained when conventional specification search is performed for the loadings by means of modification indices starting from Model 1. The fit of Model 6 ($\chi^2_{SB}(346)= 1025.20$, $p < .001$; RMSEA = .058; CFI = .917; SRMR = .037) is similar to the fit of Model 5. The quality of buffered simple structure according to Equation 6 was 4.337 revealing that no buffered simple structure was obtained. As in the population example presented above, the specification search results in fewer, but larger non-salient loadings, so that the loading pattern of Model 6 is not more parsimonious than the loading pattern of Model 5. Moreover, two Extraversion facet scales (e3, e5) did not have their largest loading on the Extraversion factor in Model 6, although they had their largest loading on the Extraversion factor in Model 5. Thus, the buffered simple structure was more compatible with the theoretical model of the questionnaire than the ICM with subsequent specification search. Overall, the example illustrates that the intended five-factor structure of the NEO-PI-R facet scales could be replicated with an acceptable model-fit when a buffered simple-structure was specified, whereas problems with the replication occurred with the ICM (bad fit) and when the ICM was combined with specification search (facets loading on unintended factors).

## 4  Discussion

Constraints for the specification of factor loadings in CFA was proposed. The constraints are based on simple structure, but they allow for non-zero secondary loadings. The motivation for proposing this criterion was that the ICM is often overly restrictive. ESEM and BSEM are also interesting alternatives to the ICM, but they have different purposes than the criterion proposed here. The criterion proposed here is based on Cattell and Tsujioka's (1964) ideas concerning 'buffered scales'. The main idea is to minimize the weighted sum of secondary or non-salient loadings. The weighted sum of non-salient loadings can be zero when the absolute size of the non-salient loadings is zero, which corresponds to the ICM. However, a zero weighted sum can also be obtained when the absolute size of non-salient loadings is substantial and when the positive and negative non-salient weighted loadings are balanced out. Since this criterion contains the ICM as a special case and since it is based on the idea of buffered unwanted variance, it was named 'buffered simple structure'.

   It was shown that buffered simple structure implies that for each block of variables with salient loadings, the corresponding sum of non-salient loadings weighted by the salient loadings should be zero. It was shown that it is possible to specify buffered simple structure by means of model constraints in CFA models. Two methods for the specification of buffered simple structure by means of CFA have been proposed. The first method uses a model constraint that avoids that buffered simple structure is reached through minimizing the absolute size of the salient loadings. This procedure is based on a single model (one-step procedure; Appendix A). The second method is based on the salient loadings and interfactor-correlations estimated from an initial ICM that are entered into a model constraint for subsequent models (Appendix B).

Thus, when an initial ICM does not fit to the data, it is possible to estimate a buffered simple structure model with model constraints based on the salient loadings and interfactor-correlations of the ICM. This illustrates that buffered simple structure can be conceived as an alternative to the ICM with subsequent specification searches.

Estimating buffered simple structure by means of model constraints is different from ESEM when the interfactor-correlations are fixed in that hypotheses are tested. In contrast, when the interfactor-correlations are freely estimated, buffered simple structure corresponds to the ESEM approach as well as to factor rotation, so that all rotational positions of the factors are equivalent in terms of model fit. ESEM might be an interesting procedure for several applications, but it is also possible that researchers prefer to remain in the context of model testing, which is possible when the interfactor-correlations are fixed and when model constraints for buffered simple structure are specified. Another alternative to buffered simple structure is BSEM. However, BSEM results in a more realistic estimation of ICM parameters, it does not address the problem that the ICM can be regarded as an overly restrictive model in the population. It may, of course, be conceived to combine BSEM with less restrictive models. However, this would be an issue for further research.

A population loading matrix with buffered simple structure was used in order to show that a perfect buffered simple structure can be found by means of the one-step procedure as well as by means of an initial ICM followed by two additional CFA-models (multi-step procedure). It was shown for the multi-step procedure that a wrong assignment of variables on the model-constraint will lead to model rejection when the factor inter-correlations are fixed to the values of the initial ICM.

It was also illustrated in the example that the strategy of estimating an ICM and to perform subsequent specification search for the loadings can be misleading when the data correspond to a buffered simple structure, because differences between loadings are induced where loadings were equal in the population. Moreover, combining ICM with specification searches leads to a mix of the confirmatory approach with an exploratory approach. This might, of course, be reasonable. However, specifying a buffered simple structure implies a priori hypotheses for the complete loading matrix (including salient and non-salient loadings) and is therefore more consistent with CFA.

A simulation study was performed in order to compare the precision of estimated population loadings and interfactor-correlations for the ICM and buffered simple structure models. It was investigated by means of the simulation study whether buffered simple structure models can be used in order to estimate salient loadings as well as interfactor-correlations when the data samples are drawn from a population corresponding to the ICM. It was found that buffered simple structure model estimates had a similar error as the ICM estimates when the population model was an ICM and when the sample size was at least moderate ($n = 300$). However, for the combination of small samples with small salient loadings and non-zero interfactor-correlations, the salient loadings of a population ICM were more precisely identified by means of an ICM. It is therefore recommended to estimate an initial ICM under these conditions and to enter the loading estimates into subsequent model constraints (see Table 1; Appendix B) under these conditions.

An empirical data set based on the facet scales of the NEO-PI-R was analyzed in order to demonstrate the CFA-specification of a buffered simple structure and to compare the results with the ICM, and ICM combined with specification search. It was found that the ICM had an inacceptable model fit and that the intended structure of the facet scales could not be replicated completely in an ICM combined with specification search, whereas it could be replicated more convincingly in the CFA-model based on buffered simple structure. It seems that forcing a maximum of non-salient loadings to be nearly zero, which is usually an aim of an ICM with subsequent specification searches, results in a substantial increase of some non-salient loadings. In contrast, buffered simple structure tends to distribute the non-salient loadings with different sign more equally across the factors.

To sum up, buffered simple structure might be regarded as a way of providing some mild constraints on the loadings as a sort of compromise between the overly restrictive CFA-ICM and could be more convincing than ICM with subsequent specification search. Accordingly, the principle of balancing non-salient loadings with positive and negative sign could complement the principle of forcing non-salient loadings to be zero. This principle of buffered simple structure could be of interest for further research, because it takes into account that non-salient loadings may not vanish, even for correlated factors. It should, however, be noted that, if the data correspond to an ICM, the ICM will also be found by means of a buffered simple structure model.

## Appendix A

```
TITLE:  POPULATION EXAMPLE, DIRECT
SPECIFICATION OF BUFFERED SIMPLE STRUCTURE (ONE-STEP PROCEDURE);

DATA:
FILE IS Example_corr.dat;
TYPE IS FULLCORR;
NOBSERVATIONS ARE 500;

VARIABLE:
NAMES ARE x1-x18;
USEVARIABLES ARE x1-x18;

ANALYSIS: TYPE = GENERAL;
ESTIMATOR=ML;

MODEL:
F1 by x1* x2-x6 (sf1_1-sf1_6);
F2 by x7* x8-x12 (sf2_7-sf2_12);
F3 by x13* x14-x18 (sf3_13-sf3_18);

F1 by x7*  (nf1_7);
F1 by x8*  (nf1_8);
F1 by x9*  (nf1_9);
F1 by x10* (nf1_10);
F1 by x11* (nf1_11);
F1 by x12* (nf1_12);
F1 by x13* (nf1_13);
F1 by x14* (nf1_14);
F1 by x15* (nf1_15);
F1 by x16* (nf1_16);
F1 by x17* (nf1_17);
F1 by x18* (nf1_18);

F2 by x1*  (nf2_1);
F2 by x2*  (nf2_2);
F2 by x3*  (nf2_3);
F2 by x4*  (nf2_4);
F2 by x5*  (nf2_5);
F2 by x6*  (nf2_6);
F2 by x13* (nf2_13);
F2 by x14* (nf2_14);
F2 by x15* (nf2_15);
F2 by x16* (nf2_16);
F2 by x17* (nf2_17);
F2 by x18* (nf2_18);

F3 by x1*  (nf3_1);
F3 by x2*  (nf3_2);
F3 by x3*  (nf3_3);
F3 by x4*  (nf3_4);
F3 by x5*  (nf3_5);
F3 by x6*  (nf3_6);
F3 by x7*  (nf3_7);
F3 by x8*  (nf3_8);
F3 by x9*  (nf3_9);
F3 by x10* (nf3_10);
F3 by x11* (nf3_11);
F3 by x12* (nf3_12);

F1@1; F2@1; F3@1;

MODEL CONSTRAINT:

0 = ( (1+sf1_1**2)*nf2_1 + (1+ sf1_1**2)*nf2_2 + (1+sf1_1**2)*nf2_3
    + (1+sf1_4**2)*nf2_4 + (1+ sf1_5**2)*nf2_5 + (1+sf1_6**2)*nf2_6 )**2;
```

```
0 = ( (1+sf1_1**2)*nf3_1 + (1+ sf1_1**2)*nf3_2 + (1+sf1_1**2)*nf3_3
    + (1+sf1_4**2)*nf3_4 + (1+ sf1_5**2)*nf3_5 + (1+sf1_6**2)*nf3_6 )**2;

0 = ( (1+sf2_7**2)*nf1_7 + (1+sf2_8**2)*nf1_8 + (1+sf2_9**2)*nf1_9
    + (1+sf2_10**2)*nf1_10 + (1+sf2_11**2)*nf1_11 + (1+sf2_12**2)*nf1_12 )**2;

0 = ( (1+sf2_7**2)*nf3_7 + (1+sf2_8**2)*nf3_8 + (1+sf2_9**2)*nf3_9
    + (1+sf2_10**2)*nf3_10 + (1+sf2_11**2)*nf3_11 + (1+sf2_12**2)*nf3_12 )**2;

0 = ( (1+sf3_13**2)*nf1_13 + (1+sf3_14**2)*nf1_14 + (1+sf3_15**2)*nf1_15
    + (1+sf3_16**2)*nf1_16 + (1+sf3_17**2)*nf1_17 + (1+sf3_18**2)*nf1_18 )**2;

0 = ( (1+sf3_13**2)*nf2_13 + (1+sf3_14**2)*nf2_14 + (1+sf3_15**2)*nf2_15
    + (1+sf3_16**2)*nf2_16 + (1+sf3_17**2)*nf2_17 + (1+sf3_18**2)*nf2_18 )**2;

OUTPUT:
```

## Appendix B

```
TITLE:  POPULATION EXAMPLE, TABLE 2, MODEL 3;

DATA:
FILE IS Example_corr.dat;
TYPE IS FULLCORR;
NOBSERVATIONS ARE 500;

VARIABLE:
    NAMES ARE x1-x18;
    USEVARIABLES ARE x1-x18;
ANALYSIS:
TYPE = GENERAL;
ESTIMATOR=ML;

MODEL:
F1 by x1* x2-x6;
F2 by x7* x8-x12;
F3 by x13* x14-x18;

F1 by x7*   (nf1_7);
F1 by x8*   (nf1_8);
F1 by x9*   (nf1_9);
F1 by x10*  (nf1_10);
F1 by x11*  (nf1_11);
F1 by x12*  (nf1_12);
F1 by x13*  (nf1_13);
F1 by x14*  (nf1_14);
F1 by x15*  (nf1_15);
F1 by x16*  (nf1_16);
F1 by x17*  (nf1_17);
F1 by x18*  (nf1_18);

F2 by x1*   (nf2_1);
F2 by x2*   (nf2_2);
F2 by x3*   (nf2_3);
F2 by x4*   (nf2_4);
F2 by x5*   (nf2_5);
F2 by x6*   (nf2_6);
F2 by x13*  (nf2_13);
F2 by x14*  (nf2_14);
F2 by x15*  (nf2_15);
F2 by x16*  (nf2_16);
F2 by x17*  (nf2_17);
F2 by x18*  (nf2_18);
```

```
F3 by x1*   (nf3_1);
F3 by x2*   (nf3_2);
F3 by x3*   (nf3_3);
F3 by x4*   (nf3_4);
F3 by x5*   (nf3_5);
F3 by x6*   (nf3_6);
F3 by x7*   (nf3_7);
F3 by x8*   (nf3_8);
F3 by x9*   (nf3_9);
F3 by x10*  (nf3_10);
F3 by x11*  (nf3_11);
F3 by x12*  (nf3_12);

F1@1; F2@1; F3@1;

F1 with F2@0.304;
F1 with F3@0.304;
F2 with F3@0.304;

MODEL CONSTRAINT:

0 = 0.600*nf2_1+ 0.600*nf2_2+ 0.600*nf2_3+
    0.600*nf2_4+ 0.600*nf2_5+ 0.600*nf2_6;

0 = 0.600*nf3_1+ 0.600*nf3_2+ 0.600*nf3_3+
    0.600*nf3_4+ 0.600*nf3_5+ 0.600*nf3_6;

0 = 0.600*nf1_7+ 0.600*nf1_8+ 0.600*nf1_9+
    0.600*nf1_10+ 0.600*nf1_11+ 0.600*nf1_12;

0 = 0.600*nf3_7+ 0.600*nf3_8+ 0.600*nf3_9+
    0.600*nf3_10+ 0.600*nf3_11+ 0.600*nf3_12;

0 = 0.600*nf1_13+ 0.600*nf1_14+ 0.600*nf1_15+
    0.600*nf1_16+ 0.600*nf1_17+ 0.600*nf1_18;

0 = 0.600*nf2_13+ 0.600*nf2_14+ 0.600*nf2_15+
    0.600*nf2_16+ 0.600*nf2_17+ 0.600*nf2_18;

OUTPUT:
```